# On-sample water content measurement for a complete local monitoring in triaxial testing of unsaturated soils

J. A. Muñoz-Castelblanco[1,2], P. Delage[1], J. M. Pereira[1] and Y. J. Cui[1]

[1] Ecole des Ponts ParisTech, UR Navier/CERMES, Université Paris-Est
6-8 av. B. Pascal, F 77455 Marne la Vallee cedex 2

[2] Now in Cathie Associates SARL
1 rue de Craiova, 92000 Nanterre

**ABSTRACT**
To provide a complete local monitoring of the state of an unsaturated soil sample during triaxial testing, a local water content measurement device was adapted to a triaxial device comprising the measurement of local displacements (Hall effect transducers) and suction (High capacity transducer). Water content was locally monitored by means of a resistivity probe. The water content/resistivity calibration curves of an intact natural unsaturated loess from Northern France extracted by block sampling at two depths (1 and 3.3 m) were carefully determined, showing good accuracy and repeatability. The validity of two models giving the resistivity of unsaturated soils with respect to their water content was examined.

The first triaxial tests carried out with this device in the range of in-situ stresses gave satisfactory results, with however some effects of the applied cell stress on the water content measurements. Some preliminary behaviour characteristics of the natural unsaturated loess, a material rarely tested up to now in the literature, were evidenced. Also, the yield stresses appeared significantly higher than the in-situ stresses, confirming the combined effect of partial saturation and bonding.

## INTRODUCTION

Following the pioneering work of Bishop and Donald (1961), various suction controlled triaxial devices have been developed to test unsaturated soils (**Table 1**). Most systems are based on the axis translation technique whereas the osmotic technique (Delage *et al*. 1987, Cui and Delage 1996) and the vapour control technique (Blatz and Graham 2000, Chávez *et al*. 2009) have also been used. Triaxial devices with internal suction measurement have also been developed using psychrometers (Verbrugge 1978, Tang *et al*. 2002 and Thom *et al*. 2008) or high capacity tensiometers (Colmenares and Ridley 2002, Meilani *et al*. 2002 and Jotisankasa *et al*. 2007).

In unsaturated soils, changes in degree of saturation are usually derived from the changes in both sample volume and water content. Local strain measurements in unsaturated triaxial soil testing should hence preferably be coupled with local measurements of water content. In this paper, a new system in which local measurements including local strains, suction and water content is presented, together with some results obtained on an unsaturated natural loess from Northern France.

## TESTED LOESS

The soil tested is an intact natural loess from a site about 100 km north of Paris. In this area, loess deposits were formed under periglacial conditions during the Quaternary period by the aeolian transport of silt particles eroded by a constant North West wind (Antoine 2002, 2003), resulting in loess deposits characterized by a relative homogeneity, a low plasticity, a high porosity and a loose microstructure. These features explain the loess susceptibility to collapse when exposed to intense rainfall or accidental water leaking (Cui *et al*. 2004, Delage *et al.*, 2005, Yang *et al.*, 2008, Munoz-Castelblanco et al. 2011) as well as its liquefaction behaviour under cyclic loading (Karam *et al*. 2009).

Quality specimens were trimmed from intact cubic block samples (300 mm side) extracted at 1m and 3.3 m depths. The geotechnical properties of the samples are presented in Table 2 and the grain size distribution curves are shown in Figure 1. Both samples, classified as CL in Casagrande's classification, have a relatively low clay fraction and low plasticity index (9). The dominant clay minerals are kaolinite, illite and interstratified illite-smectite (Karam 2006). The carbonate content values are close (6% at 1m and 5% at 3.3 m). Due to a higher void ratio and a lower degree of saturation, the 1 m depth sample exhibits a collapse potential slightly higher than that of the 3.3 m depth sample.

## TRIAXIAL APPARATUS

### General instrumentation

A triaxial cell (Figure 2) was designed to accommodate samples of 50 mm diameter and 100 mm height so as to achieve better accuracy in local measurements. Hall-effect transducers were used for the measurements of both radial and axial displacements. The axial displacement was also measured using an external LVDT transducer. As seen in Figure 3 (left), suction was locally measured at mid-height of the sample (Meilani *et al.* 2002, Colmenares and Ridley 2002, Jotisankasa *et al*. 2009) using a miniature 500 kPa in-house constructed high capacity tensiometer (HCT, Ridley and Burland 1993, Cui *et al.* 2008) of diameter 5 mm (Chiu *et al.* 2002). Cycles of cavitation-saturation were carried out to improve the HCT performance (Tarantino and Mongiovi 2001, Cui *et al.* 2010). Water content was monitored using a specially designed new electrical resistivity probe of diameter 11 mm. The tests were carried out using strain controlled method.

### Local Measurement of water content

To investigate in the laboratory the relationship between water content and electrical resistivity, Gupta and Hanks (1972) and Rhoades *et al.* (1976) tested compacted specimens using circular four-probe resistivity cells, a device also used by Kalinski and Kelly (1993). Other resistivity measurements were made by Fowles (1980) on compacted specimens, McCarter (1984) on remoulded clays, Fukue *et al.* (1999) on remoulded and natural clays and Chen *et al.* (2007) on expansive soils. As quoted by Kalinski and Kelly (1993), the resistivity of saturated soils depends on the particle size distribution, mineralogy, specific clay surface, porosity, pore size distribution, connectivity of pores, water content, salt concentration and temperature.

In unsaturated soils, the electrical resistance depends on that of solids $R_s$, of air $R_a$ and of water $R_w$. Whereas the air phase is an electrical insulator, water has an electrical conductivity significantly higher than solids and the electrical current flows preferably through it. In clayey soils, clay water interactions tend to reduce the mobility of the adsorbed water molecules resulting in a smaller resistivity at higher water content when more free water is available. This reduction seems however negligible for soils of low specific surface area ($S_s < 50$ m$^2$/g, Fukue *et al.* 1999). Thus, it should not be significant in the loess studied here.

For saturated rocks, Archie (1942) proposed a simple empirical model relating the rock electrical resistivity $\rho$ and the pore water electrical resistivity $\rho_w$ to the porosity *n*, as follows:

$$\frac{\rho}{\rho_w} = (n)^{-a} \tag{1}$$

Archie's second law extends this model to unsaturated state by introducing the degree of saturation $S_r$ (see e.g. Guéguen and Palciauskas, 1994):

$$\frac{\rho}{\rho_w} = (n)^{-a} (S_r)^{-b} \tag{2}$$

where *a* and *b* are soil constants.

Fukue *et al.* (1999) proposed a more sophisticated model accounting for the combined effects of the serial and parallel transmission of the electric current in the three phases (air, water and solids). They defined a structural coefficient *F* to separate the parallel flux (related to $1 - F$ and mainly occurring in water) and the serial one (related to *F* and influenced by the insulating properties of solids and air), giving the following expression of the electric resistivity $\rho$ through a cylinder of radius *r*:

$$\rho = \frac{\pi r}{w G_s n} \Gamma \; ; \; \Gamma = \frac{\rho_w}{(1-F)} \tag{3}$$

where *w* is the gravimetric water content, $G_s$ is the specific solid density and *n* the soil porosity. The coefficient *F* has the dimension of a length and depends on the structure of the soil. The quantity $\Gamma$ can be related to the soil state.

In this study, a small-sized electric resistivity probe (11 mm in diameter) was developed to measure the water content at the mid-height of the specimen. It was inspired from the concept of concentric and surface probe developed by Maryniak *et al.* (2003). The probe is composed of four circular electrodes of diameter 1.5 mm disposed in a squared-grid (inter electrodes distance of 6 mm) as presented in Figure 4. A hydrophobic and dielectric matrix (Araldite epoxy resin) was used to accommodate the electrodes with proper electric insulation). Figure 3 shows how the resistivity probe was fixed on the triaxial sample opposite to the suction probe. The electrodes connection is presented in Figure 5. Two input electrodes are supplied by a voltage source of 10 V. The current passes through the soil (characterized by its resistance $R_s$) and two output electrodes receives the output signal.

In a circuit in parallel:

$$\frac{1}{R} = \frac{1}{R_{s1}} + \frac{1}{R_{s2}} ; \tag{4}$$

giving $R_s = 2R$ because $R_{s1} = R_{s2} = R_s$

The shape of the current lines between the input and output electrodes is related to the geometry and boundary of the problem. The electrical resistivity is given as:

$$\rho = \frac{R_s A_e}{L} \tag{5}$$

in which $R_s$ is the measured soil electric resistance, $A_e$ is the smallest electrode surface and $L$ is the shorter distance between each pair of electrodes.

When an electrical signal comes out from the input electrode, an instantaneous output response is provided by the closest electrodes. Once the initial current peak appears, the electric signal decreases due to diffusion through the soil. To avoid any parasite effect due to diffusion, the electric current was imposed during periods shorter than 20 s, every 6 minutes. This period was determined after running a calibration test to determine the time needed to disperse all the electrical charge after each current application. This operation was needed to avoid any polarization in the soil and any effect in the tensiometer. The time taken allowing half the dispersion to occur is termed relaxation time $\tau$ and is related to the characteristic frequency $f_0$ (Debye 1929, Mitchell and Arulanandan 1968) according to:

$$\tau = \frac{1}{2\pi f_0} \qquad (6)$$

Electric dispersion causes several types of polarization in soils (Fam and Santamarina 1995): electronic, molecular, Maxwell-Wagner and macroscopic polarization, each of which having a range of characteristic frequency. The lower limit of the characteristic frequency of the macroscopic polarization is 0.001 Hz corresponding to a maximum relaxation time of 2.65 min, a value that coincides with the half period of electrical dispersion obtained experimentally (3 min).

A calibration was previously performed to characterize the relationship between the water content and the soil resistivity. Five triaxial loess specimens (three from 1m depth and three from 3 m depth) of height 100 mm and diameter 50 mm were submitted to controlled wetting and drying processes while measuring their electrical resistivity at mid-height with the gauge presented in Figure 4. To ensure water content homogeneity, drying was performed by allowing evaporation of the sample water under laboratory conditions for periods of time comprised between 10 and 24 hours. Homogeneous wetting was achieved by carefully adding small drops of water with a syringe to the soil sample through two pieces of filter paper placed on top and bottom. Water content changes at equilibrium were controlled by weighing to an accuracy of 1/1000 g.

The calibration data obtained on samples from 1 m and 3.3 m depth along both the wetting and drying paths are shown in Figure 6. For each depth, good compatibility between the results of the different specimens is observed. At 1 m depth, the soil resistivity increases from 7 Ωm to 350 Ωm when the volumetric water content θ decreases from 36 to 6 % (w decreasing from 25% to 4%). The slope of the curve indicates that a reasonable estimate can be made with volumetric water content values higher than 7% (w > 5%), whereas it becomes less accurate in drier states with resistivity values higher than 100 Ωm. The soil resistivity at natural water content (ρ = 30 Ωm at w = 14.4%) is also presented in this figure together with that of water (4 Ωm). In samples from 3.3 m depth, the resistivity changes between 9 Ωm to 750 Ωm when θ decreases from 31% to 9% (w between 22% and 7%) with a smaller slope allowing good determination when 13% < θ < 31% (8% < w < 22%). The lower initial resistivity of the deeper sample (18 Ωm at 3.3 m compared to 30 Ωm at 1 m) is related to its smaller void ratio and higher initial degree of saturation ($e$ = 0.60 at 3.3 m compared to 0.84 at 1 m; $S_r$ = 76% - 80% at 3.3 m compared to 46% at 1 m).

The data from both soils are also presented in a plot of $\rho/\rho_w$ versus $S_r$ in Figure 7 and compared to the extended Archie's law and to the model proposed by Fukue *et al.* (1999). Archie's curves agree reasonably well in the range of degree of saturation from 15% and 70% for the 1m depth specimen and in the range of degree of saturation from 30% and 70% for the 3.3 m depth specimen. Given their concave shape, Fukue's curves better agree with experimental data than the Archie's model, especially for degrees of saturation between 15% and 75% for the 1 m depth specimens and between 25% and 50% for the 3.3 m depth specimens. Fukue's values F are 0.94 m for the 1 m depth specimen and 0.97 m for the 3.3 m depth specimen, corresponding to Γ values of 297 Ω and 366 Ω, respectively. There is no evidence of a specific or quantitative meaning of the *F* value. Fukue *et al.* (1999) stated that parameter Γ is an indicator of the soil state and that its values close to or higher than 300 Ω indicate that specimens are undisturbed. Thus, the values fitted in this study show the good quality of the extraction and sampling procedures followed.

The stability of the resistivity probe response with respect to time was checked on the 1 m depth soil using a sample of height 20 mm and diameter 70 mm (Figure 8). The overall sample water content was kept constant at 14.1% by enveloping the sample in a plastic film and a slurry layer of the same loess was put over the probe to avoid probe cavitation during its installation. The initial measured water content of 26.7% corresponds to that of the slurry. As seen in Figure 8 (top), the water content is observed to linearly decrease owing to water transfer from the slurry into the sample with equilibration reached after 16 h. Once the soil water content is attained, little further variations are observed with a water content stabilised at 14.15±0.03%, in good agreement with the overall water content determined by weighing. Note that the amount of water infiltrated from the slurry is sufficiently small to not significantly affect the sample water content. The data of Figure 8 indeed shows an excellent stability of the water content measurement.

**TESTING PROCEDURES AND RESULTS**

After trimming, the dimensions and weight of each sample were taken with a calliper to an accuracy of 1/1000 mm. At the soil natural state, the water content was about 14% and suction around 45 kPa. To increase the water content, the samples were enveloped with filter paper and carefully sprayed. Drying was made by controlled evaporation as mentioned before. A period of equilibration of at least 24 h was waited for to ensure water content homogeneity. Target suctions were achieved based on the water retention curve of the tested loess, obtained in a previous work by Muñoz-Castelblanco *et al.* (2011). The sealing of all local sensors was done with silicone to ensure water tightness without increasing the membrane stiffness. Silicone oil was used as confining fluid.

**Constant water mass shearing, 3.3 m depth sample**

To investigate the reliability of the device, two similar tests were conducted on two specimens trimmed from the same block extracted at 3.3 m. The transient suction equilibration phase with water exchange between the tensiometer, the slurry and the sample is presented in Figure 9 for both samples. It lasts about 20 minutes in both cases, a period of time quite similar to that observed by Meilani *et al.* (2002) in the same conditions. Stabilisation is afterwards observed at the same suction for both specimens ($s_0$ = 50 kPa). This shows the validity of the suction measurements and the good homogeneity of suction distribution within the block from which the two specimens were trimmed.

The data of the two triaxial tests performed are presented in Figure 10 in terms of changes in deviator stress *q*, volumetric $\varepsilon_v$ and radial $\varepsilon_3$ strain (local measurements from the Hall effects transducers), suction *s* and water content *w* with respect to the local axial strain $\varepsilon_1$ on the right side of the figure. The same data are presented with respect to the changes in net mean stress $p - u_a$ on the left side of the figure.

The stress path followed is indicated on the left top of the figure. The samples were firstly brought back to the in-situ net stress ($\sigma_v - u_a = 65$ kPa at a depth of 3.3 m) following a $K_0$ path (no lateral strain allowed, see the $\varepsilon_3$ versus $p - u_a$ diagram in the Figure). A comparable stress path on unsaturated soils has recently been followed by Boyd and Sivakumar (2011). Note that due to the important concern related to the validity of the effective stress in unsaturated soils, $K_0$ is defined here in terms of net stresses, as follows:

$$K_0 = \frac{\sigma_3 - u_a}{\sigma_1 - u_a} \tag{7}$$

in which $\sigma_1$ is the axial stress, $\sigma_3$ is the confining pressure and $u_a$ the air pressure (atmospheric pressure). The test provided a value of the lateral net stress of 25 kPa corresponding to a $K_0$ value of 0.38. The stress paths of the two tests appear quite close in all the $q/\varepsilon_1$, $\varepsilon_v/\varepsilon_1$ and $\varepsilon_3/\varepsilon_1$ plots. Note also that the volume change is quite small along the $K_0$ path.

A standard test with constant cell pressure ($\sigma_3 = 25$ kPa) was carried out starting from the in-situ net stress conditions. The $q/\varepsilon_1$ stress-strain curves obtained exhibit in both cases a typical shape and provide estimated yield stress values that are also reported in the other graphs. Note that the yield stress defined here delimits a pseudo-elastic zone (Moulin 1989, Maâtouk et al. 1995, Cui and Delage 1996). It is not the true elastic limit under which strictly no irrecoverable strain occurs. Indeed, deeper examination of the stiffness (secant modulus $E_s$) variations with axial strain shows that this true elastic domain with no variation of stiffness ($E_s$ is at its maximum value) is in the range of very small strain ($\varepsilon_1 < 0.01\%$). The values of yield stress determined are close ($q = 141$ kPa and 148 kPa and ($p - u_a$) = 72 kPa and 74 kPa, for Tests 1 and 2 respectively), they are significantly larger than the in-situ stress that is also presented in the plot. Since this Quaternary aeolian deposit has never been covered by any other soil layer since the last glaciation, overconsolidation effects cannot be considered here. In loess, it is well known that the process of dissolution and reprecipitation of primary carbonate minerals provide some carbonate bonding between the silt particles (Pecsi 1991), explaining the structure effects that provide a higher yield stress than the in-situ stress. These effects are strengthened by the suction hardening due to desaturation (Gens and Nova 1993, Cui and Delage 1996) which also increases the apparent yield stress values. Little volume change is observed before yield whereas compressive strain is afterwards progressively developing, in accordance with the high void ratio, the low initial degree of saturation and the collapsible behaviour of the loess.

The local changes in water content are presented together with the initial and final global values estimated by weighing, showing that little change occurred during the test. In both cases, the initial measured values (under zero stress, as during the calibration) present a good correspondence with the gravimetric water contents, whereas an increase of around 0.3% is observed between the initial and yield states (with a mean stress close to 75 kPa). These slight changes are thought to result from the application of the mean stress on the gauge. Indeed, the contact between the electrode and the soil can be improved by the application of the confining stress.

The measurements are afterwards relatively stable in the case of test 1 within a range of 0.25% whereas a global decrease is observed in test 2. The final measured value is larger (0.25%) than the gravimetric one in test 1 and smaller (0.5%) in test 2. Test 2 is obviously not fully satisfactory and the measured decrease could be due to temperature changes (that were not monitored) during the test.

The local changes in suction observed in the $s/(p – u_a)$ plot show an initial decrease of 9-14 kPa from the initial values (from 51 to 42 kPa and 54 to 40 for samples 1 and 2 respectively) that occurs when bringing the samples back to its in-situ conditions under $K_0$ conditions at constant water content. No significant volume change is observed during this phase. Actually, in both tests, these suction changes appear to be coupled with that of water content, probably indicating here also an effect of the improvement of the soil/gauge contact with the increase in confining stress. Afterwards, small changes affect the suction values during shearing. Note that a failure plane appeared in both tests.

**Constant water mass shearing, 1 m depth sample**

A constant rate of strain shear triaxial test at constant confining stress of 8 kPa (close to the lateral in-situ stress) was performed on a 1m depth loess specimen at the initial in-situ water content ($w_i$ = 14%). Testing was performed at a constant vertical displacement rate of 10 µm/min.

The complete set of data is presented in Figure 11 in the same fashion as in Figure 10. The yield stress estimated from the $q/\varepsilon_1$ curve is reported in other plots. As previously, the yield stress ($q_y$ = 35 kPa) is far above the in-situ stress ($q$ = 20 kPa) with little volume change observed before yield whereas compressive strain is afterwards progressively developing (see the $\varepsilon_v/(p- u_a)$ diagram on left side).

The local water content initially decreases from 14% to 13.2% up to an axial strain of 0.6%, whereas the global water content slightly decreases from 14 to 13.77%. The water content is afterwards satisfactorily stable. Here also, the application of the confining stress seems to result in a decrease of water content. Conversely, suction change is not affected by the confining stress. After a slight decrease, the local suction progressively increases from 40 kPa to 60 kPa.

**DISCUSSION**

In spite of some problems apparently related to the effects of the confining pressure on the contact between the resistivity gauge and the soil specimen, the data from tests 1 and 2 in Figure 10 and from the test of Figure 11 on intact natural loess specimens from two depths (1 and 3.3 m), showed that the water content locally monitored by a new resistivity probe remained almost constant during shearing. In the first series of tests (depth 3.3 m), the initial suction change appeared to be also due to the changes in contact between the gauge and the sample during the application of the confining stress. In both tests, suction afterwards remained reasonably constant.

The progressive suction increase between 40 and 60 kPa observed under constant water content in the test at 1 m (Figure 11) is somewhat comparable to the suction changes obtained during shearing by Colmenares and Ridley (2002) on a compacted silty clay. This increase occurred during a progressive decrease in volume that can be related to the high porosity and the collapse susceptibility of the sample at 1.1 m, as evidenced by Munoz-Castelblanco et al. (2011). This trend, that obviously deserves further interpretation, could be due to significant effects of the increase in shear stress.

## CONCLUSIONS

A new triaxial cell with complete local monitoring of the state of an unsaturated specimen was designed. Besides local strain and suction measurements, the system also includes a new in-house constructed resistivity probe for the local measurement of water content. The design and the calibration of this new device were carried out on two intact natural loess specimens and the validity of two existing resistivity models was examined.

Two sets of preliminary triaxial tests at a constant confining pressure and constant water content starting from the in-situ stress conditions were conducted on loess specimens. The small water content changes that were observed during the application of the cell load evidenced some effects of the stress on the water content measurements, probably due to the improvement of the sample/gauge contact with the applied pressure. Once this adjustment occurred, most tests indicated a fairly constant value of water content during shearing. Whereas some effect of the applied cell pressure also affected the suction measurement in the series of tests at 3.3m, the response in suction in the test at 1 m provided during shearing a regular increase that is comparable with the few available published data.

The two loess specimens exhibited a fragile behaviour characterized by yield stresses significantly higher than the in-situ stresses, a trend typical of structured unsaturated soils. These data provide some first behaviour characteristics of an intact natural unsaturated soil. This kind of data appears to be quite rare in the literature, since most existing data concern artificial unsaturated samples.

## ACKNOWLEDGMENTS

The present study is part of the first author PhD work. It was supported by the European Alβan Program of high level scholarships for Latin America, scholarship N° E07D402297CO, through grants to Mr. J. Muñoz. The support of the French Railways Company SNCF is also acknowledged.

Table 1. Triaxial systems to test unsaturated soils

| Author | Soil volume changes (Devices) | Suction (control and/or measurement) | Observations |
|---|---|---|---|
| Bishop and Donald (1961) | Internal cylinder full of mercury | Axis translation (control). Air pressure applied on top | |
| Maâtouk *et al.* 1995 | Vertical external deflectometer and lateral strain monitoring device | Axis translation (control). Air pressure applied through a hole at specimen mid-height. | - |
| Wheeler and Sivakumar 1995 | Double cell system | Axis translation (control) | - |
| Cui and Delage 1996 | Double cell system | Osmotic technique (control) | - |
| Hoyos and Macari 2001 | True triaxial strain system | Axis translation (control) | True triaxial device |
| Barrera 2002 | Local strain measurement. Global measurement of volume change and water content | Axis translation (control) | |
| Aversa and Nicotera 2002 | Double cell system | Axis translation (control) | Water content measurement with double walled burettes system |
| Matsuoka *et al.* 2002 | True triaxial strain system | Axis translation (control) | - |
| Sivakumar *et al.* 2006 | Twin-cell stress path apparatus | - | - |
| Cabarkapa and Cuccovillo 2006 | LVDT, Radial strain belt | Axis translation (control) | - |
| Padilla *et al.* 2006 | Double cell system | Axis translation (control) | - |
| Cui *et al.* 2007 | Double cell system | - | Cyclic loading at different water content levels |
| Jotisankasa *et al.* 2007 | Local axial strain devices and radial strain belt | Two suction probes at two heights of the specimen (measurement) | Drying and wetting paths under loading |
| Rojas *et al.* 2008 | Double cell system | Axis translation (control) | - |
| Thom *et al.* 2008 | Hall Effect transducers | Thermocouple psychrometer (measurement) | - |
| Xu *et al.* 2008 | Hall Effect transducers | Axis translation (control) | - |
| Blatz and Graham 2000 | | Relative humidity | |
| Chávez *et al.* 2009 | Double cell system / LVDT and DDT transducers | Relative humidity control (vapour transfert) | - |

Table 2. Geotechnical characteristics of the Bapaume loess

| Sample depth | 1 m | 3.3 m Test 1 | 3.3 m Test 2 |
|---|---|---|---|
| Natural water content $w$ (%) | 14.0 | 17.9 | 17.0 |
| Natural void ratio $e$ | 0.84 | 0.60 | 0.60 |
| Dry unit mass $\rho_d$ (Mg/m$^3$) | 1.45 | 1.67 | 1.67 |
| Natural degree of saturation $S_r$ (%) | 46 | 80 | 76 |
| Natural suction (HTC) (kPa) | 40 | 48 | 49 |
| Clay fraction (% < 2 μm) | 16 | 25 | 25 |
| Plastic limit $w_p$ | 19 | 21 | 21 |
| Liquid limit $w_l$ | 28 | 30 | 30 |
| Plasticity index $I_p$ | 9 | 9 | 9 |
| Carbonate content (%) | 6 | 5 | 5 |
| In situ total vertical stress $\sigma'_{v0}$ (kPa) | 15.47 | 35.57 | 35.57 |
| Collapse (%) under $\sigma'_{v0}$ | 7.5 | 6.3 | 6.3 |

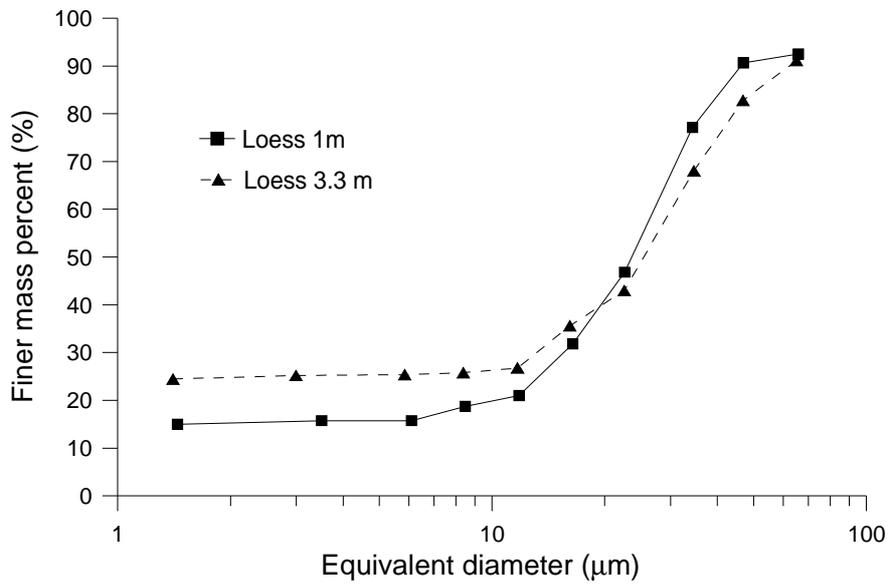

Figure 1. Grain size distribution of Bapaume loess samples from 1 and 3.3 m depths

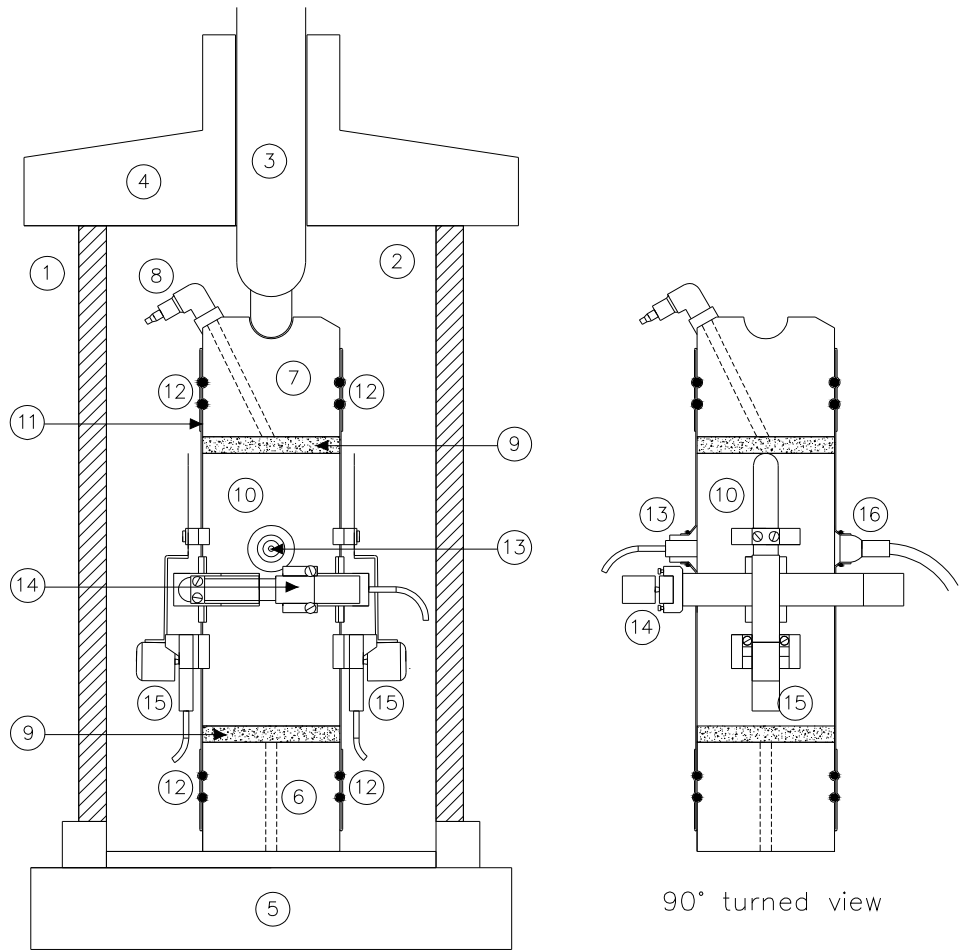

1. Perspex cylinder
2. Cell chamber (1 MPa)
3. Loading piston
4. Top plate
5. Cell base
6. Lower cap
7. Top cap
8. Top drainage
9. Porous stone
10. Soil specimen
11. Latex membrane
12. o-rings
13. Tensiometer
14. Radial strain transducer (Hall Effect)
15. Axial strain transducer (Hall Effect)
16. Resistivity probe

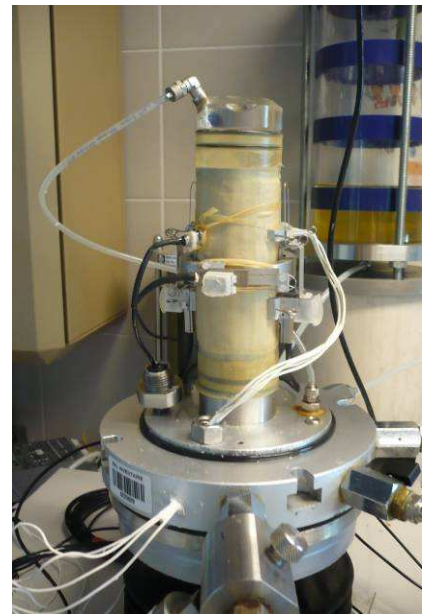

Figure 2. Modified triaxial cell

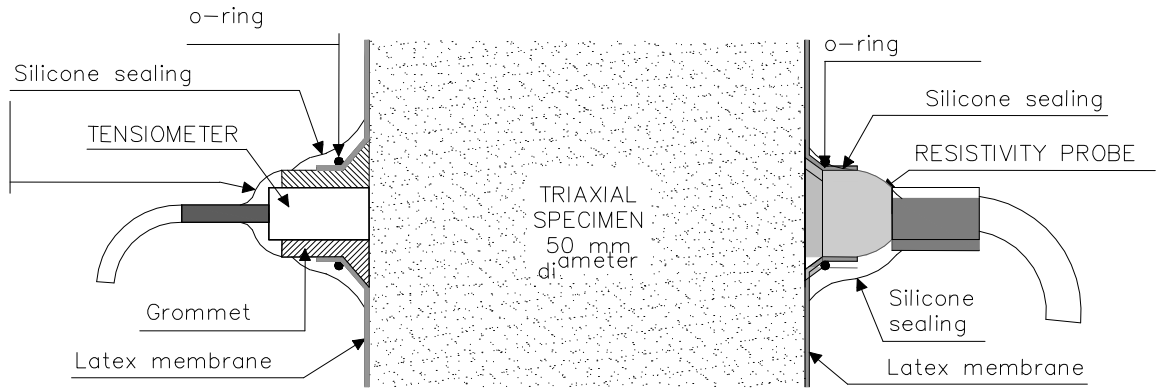

Figure 3. Triaxial specimen. Lateral view. Holding system for the tensiometer (left) and the resistivity probe (right)

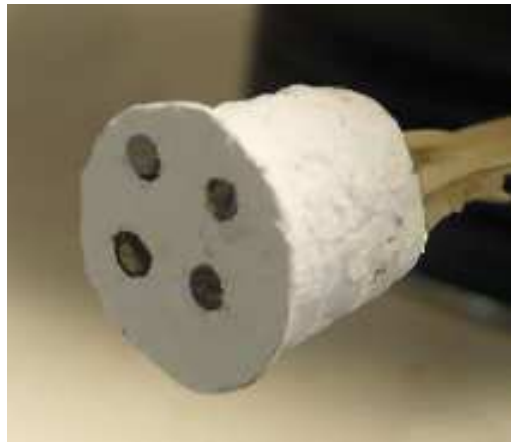

Figure 4. In house constructed resistivity probe

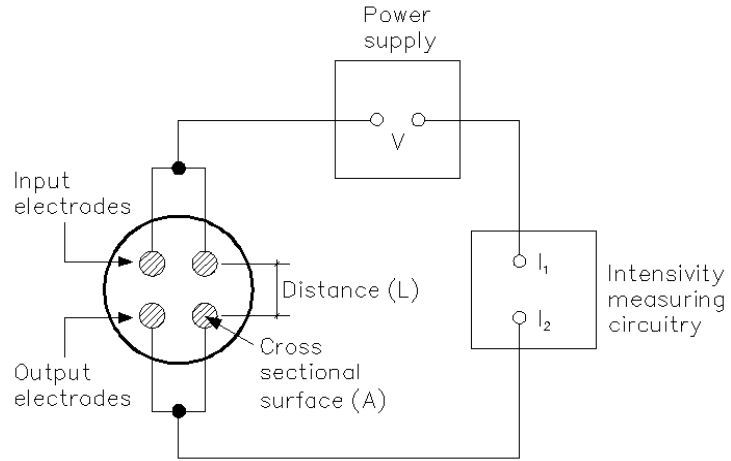

Figure 5. Electric resistivity device with four electrodes.

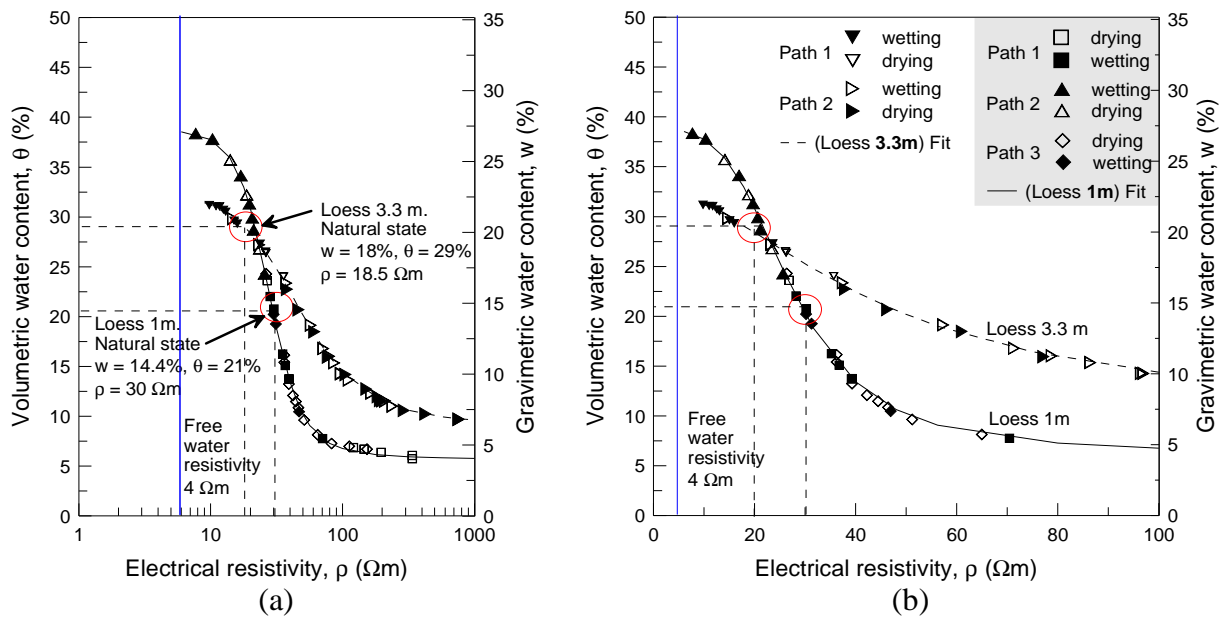

Figure 6. Calibration curves of the resistivity probe for loess at 1m depth and 3.3 m depth. Electric resistivity v.s. volumetric water content. (a) Resistivity values in log-scale. (b) Zoom in natural scale

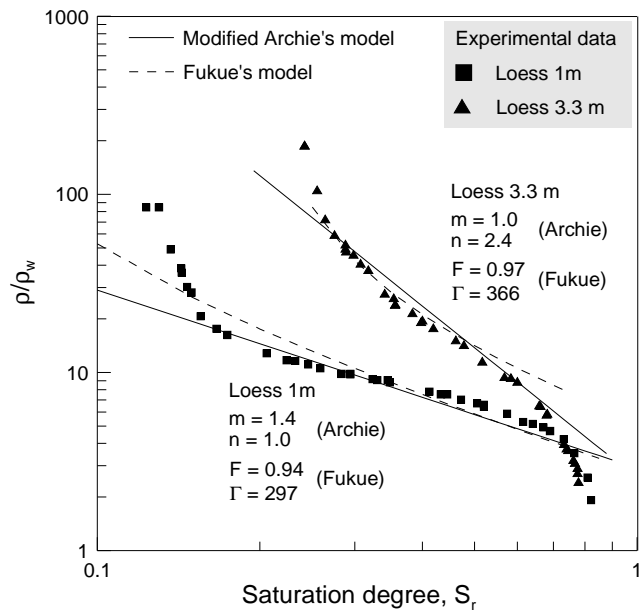

Figure 7. Resistivity data. Comparison with Archie's second law and Fukue's model

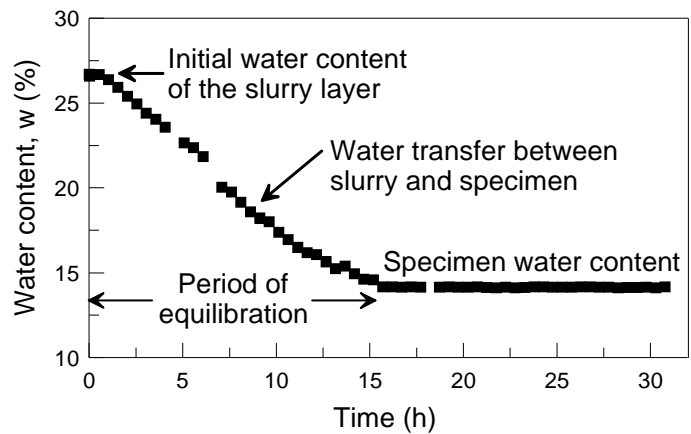

Figure 8. Response of the resistivity probe. Evolution of water content with time.

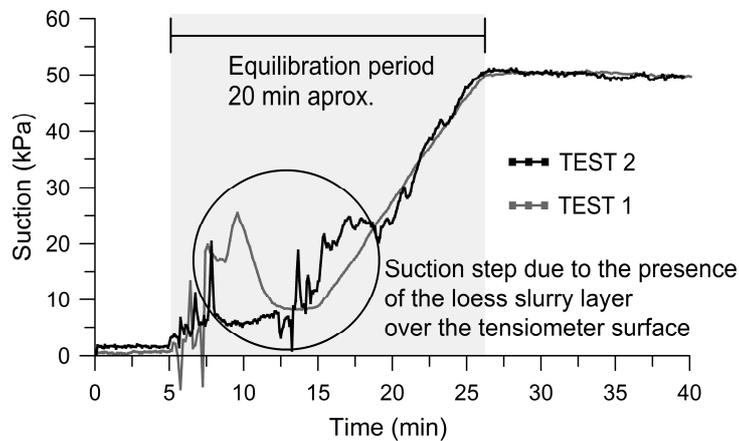

Figure 9. Suction measurement for 3.3 m loess samples during equilibration between tensiometer and sample

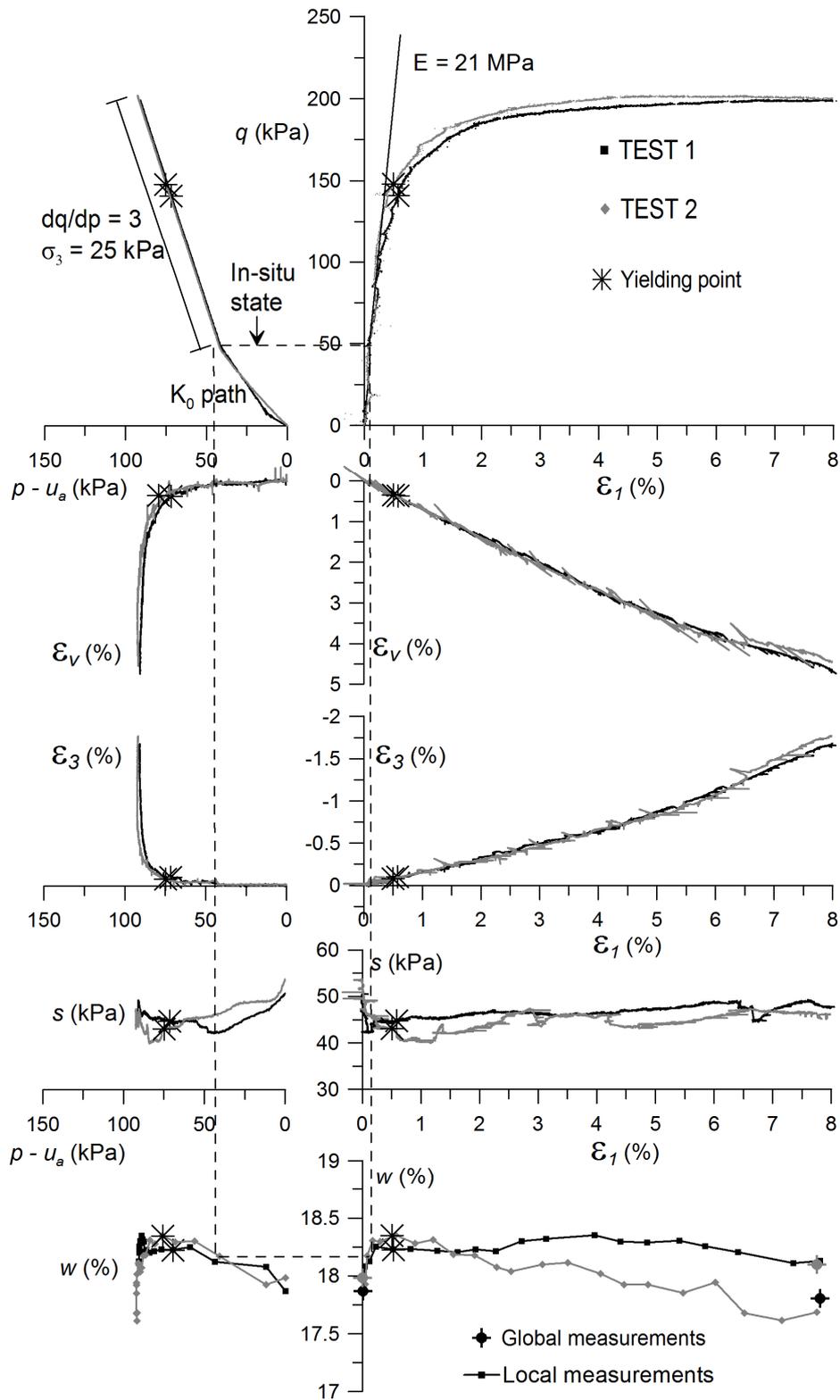

Figure 10. Triaxial tests on 3.3 m loess samples at their natural state (w ≈ 18%, s ≈ 50 kPa). Both samples are from the same block

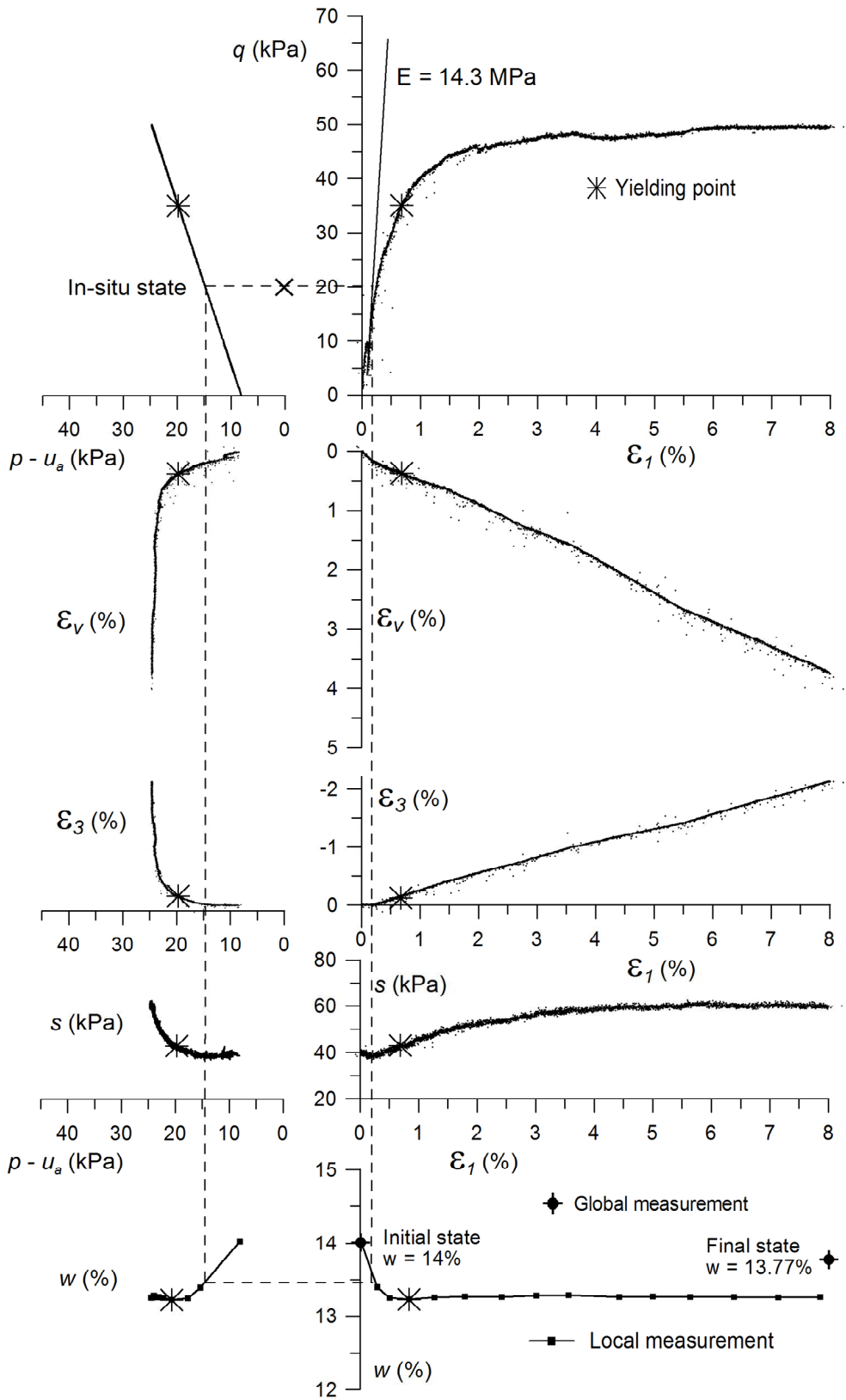

Figure 11. Triaxial tests results for 1m loess sample under constant confining pressure of 8 kPa corresponding to the natural in-situ lateral stress